# La conservazione dei documenti informatici nel contesto sanitario italiano: indagine su stato di attuazione e criticità

Maria Teresa Guaglianone, Elisa Sorrentino, Elena Cardillo, Maria Teresa Chiaravalloti, Giuseppe Alfredo Cavarretta*

**Abstract:** The realization of digital preservation in compliance with legislation is extremely important, especially in a sensitive domain like healthcare, where guaranteeing document reliability, authenticity, integrity and readability over time is essential to have an immediate return in terms of efficiency of the whole care setting. In this perspective, the present paper highlights critical issues, through detailed surveys addressed to both national health facilites and digital preservers, defining the state of the art of digital preservation practices in the Italian healthcare setting. The final aim is to identify the strategic areas that need technical and regulatory interventions, in order to offer a major boost of innovation to the domain. Results show an extremely variegated context that is not always compliant to the articulated legislation. These results will be used to integrate the new Italian *Guidelines on the creation, management and preservation of digital documents* published by the Agency of Digital Italy.

## 1. Introduzione

La conservazione dei documenti informatici, oltre a rappresentare un obbligo normativo, si rivela una pratica di fondamentale importanza poiché garantisce, nel medio e lungo periodo, il mantenimento delle caratteristiche di affidabilità, integrità, autenticità e la reperibilità dei documenti e dei fascicoli informatici così come previsto dall'art. 44 del Codice dell'Amministrazione Digitale (CAD[1]). In altri termini, la conservazione digitale a norma garantisce la possibilità di mantenere tali caratteristiche nel tempo e in contesti diversi da quello di produzione, assicurando il pieno valore legale dei documenti (rendendoli opponibili a terzi in giudizio), e la possibilità di reperire documenti che siano stati prodotti e archiviati in maniera ordinata e strutturata, attraverso l'associazione di specifici metadati[2]. Ciò è ancor più vero in un settore quale quello sanitario, in cui la disponibilità di dati e documenti è fondamentale per poter predisporre azioni adeguate soprattutto con finalità di cura, oltre che di ricerca e di governo.

In considerazione di ciò, il presente contributo mira ad analizzare lo stato dell'arte delle pratiche di conservazione in uso nel contesto nazionale e, in particolare, presso le strutture sanitarie delle diverse Regioni e Province Autonome per poter identificare eventuali criticità nell'implementazione del processo di conservazione e nell'adozione di metodologie, tecnologie e strumenti utilizzati, così da definire i prerequisiti per un corretto approccio alla conservazione digitale in ambito sanitario. Tale analisi è stata effettuata attraverso un'indagine, condotta dall'Istituto di Informatica e Telematica del Consiglio Nazionale delle Ricerche (IIT – CNR), in collaborazione con l'Agenzia per l'Italia Digitale (AgID), avente come obiettivo finale quello di redigere un allegato tecnico, specificatamente dedicato alla documentazione sanitaria, alle nuove Linee guida ex art. 71 CAD, di recente pubblicazione sul sito di AgID[3].

---

* Istituto di Informatica e Telemarica, Consiglio Nazionale delle Ricerche, Rende (CS), Italia.
maria.guaglianone@iit.cnr.it; elisa.sorrentino@iit.cnr.it; elena.cardillo@iit.cnr.it.; maria.chiaravalloti@iit.cnr.it; giuseppealfredo.cavarretta@cnr.it.

[1] Recentemente modificato dal Decreto Legislativo 26 agosto 2016 n.179, *Modifiche ed integrazioni al Codice dell'amministrazione digitale*, di cui al Decreto Legislativo 7 marzo 2005, n. 82, ai sensi dell'articolo 1 della legge 7 agosto 2015, n. 124, in materia di riorganizzazione delle amministrazioni pubbliche. (16G00192) (GU Serie Generale n.214 del 13-09-2016).

[2] È possibile associare al documento diverse tipologie di metadati: descrittivi (per l'identificazione e il recupero degli oggetti digitali), amministrativi e gestionali (per fornire informazioni utili alla gestione, all'accesso e alla conservazione delle risorse descritte), strutturali (per descrivere la composizione strutturale del documento al quale sono associati). In particolare, i metadati amministrativi e gestionali svolgono un importante ruolo all'interno del processo della conservazione digitale, poiché consentono di «documentare i processi tecnici associati alla conservazione permanente, fornire informazioni sulle condizioni ed i diritti di accesso agli oggetti digitali, certificare l'autenticità e l'integrità del contenuto, documentare la catena di custodia degli oggetti, identificarli in maniera univoca». A. Scolari, M. Pepe, M. Messina, C. Leombroni, G. Cirocchi, G. Bergamin, Appunti per la definizione di un set di metadati gestionali-amministrativi e strutturali per le risorse digitali. Gruppo di studio sugli standard e le applicazioni di metadati nei beni culturali promosso dall'ICCU, versione 0, 2002, <https://www.iccu.sbn.it/export/sites/iccu/documenti/MetaAGVZintroduzione.pdf> (ultima consultazione 16/03/2020).

## 2. Stato dell'arte e contesto normativo

Gran parte della letteratura sul tema della conservazione digitale comprende sia lavori che descrivono analisi generalizzate sull'argomento sia lavori focalizzati su un'analisi specifica del settore sanitario, ma nessuno di essi riguarda indagini recenti volte a verificare, a livello nazionale, l'effettivo stato di applicazione della normativa e di realizzazione di sistemi di conservazione a norma. La letteratura sull'argomento pone, piuttosto, l'attenzione sull'analisi della nomativa di riferimento e sulla descrizione delle diverse iniziative attuate, anche a livello internazionale, in materia di conservazione.

Un esempio di questo tipo di studi si riscontra in un recente lavoro di Pigliapoco[4], in cui viene posta l'attenzione sull'analisi e sull'armonizzazione della normativa di riferimento. Vi sono poi lavori che riguardano indagini conoscitive sulle normative e sulle linee d'azione per la conservazione delle memorie digitali in uso nei Paesi europei e in alcune importanti istituzioni internazionali, rispetto alla conservazione delle risorse digitali, che affrontano, nello specifico, questioni relative all'introduzione di interventi regolamentari, alla definizione delle responsabilità connesse alla loro produzione e mantenimento e alle attività di monitoraggio e revisione, facendo riferimento alla definizione di costi e benefici di un sistema normativo a più livelli[5].

Un'indagine simile per obiettivi alle attività condotte nel presente studio è presentata in un lavoro di Guercio[6], nel quale si indaga il quadro giuridico nazionale e le buone pratiche per la conservazione digitale.

Il presente studio, sulla scia di queste esperienze pregresse, mira ad aggiornare questo tipo di indagine con dati recenti e, soprattutto, si focalizza sulla conservazione nello specifico settore della sanità pubblica in Italia.

Da un punto di vista strettamente normativo, le già citate recenti modifiche al CAD sono il riferimento principe di settore, mentre più specificatamente rivolto all'ambito sanitario è il DPCM 178/2015 *Regolamento in materia di fascicolo sanitario elettronico,* che all'art. 23 detta "Misure di sicurezza e sistema di conservazione", evidenziando che ciascun sistema regionale di Fascicolo Sanitario Elettronico (FSE) è tenuto a predisporre la conservazione dei documenti informatici di tipo sanitario garantendo confidenzialità (riservatezza), integrità e disponibilità dei dati. Sulla scorta di quest'ultimo è stata predisposta la prima parte del lavoro di indagine su FSE e conservazione dei documenti informatici, che viene presentato nel presente contributo, mentre il D. Lgs. 179/2016 è stato poi il riferimento per approfondire gli aspetti più legati nello specifico alla documentazione sanitaria nella seconda parte del presente lavoro.

---

[3] AGENZIA PER L'ITALIA DIGITALE, *Linee guida sulla formazione, gestione e conservazione dei documenti informatici*, Release bozza 13 febbraio 2020, <https://docs.italia.it/AgID/documenti-in-consultazione/lg-documenti-informatici-docs/it/bozza/index.html>, (ultima consultazione: 16/03/2020).
[4] S. PIGLIAPOCO, *La conservazione digitale in Italia. Riflessioni su modelli, criteri e soluzioni*, in «JLIS.it», January, vol. 10, n. 1, 2019, pp. 1–11.
[5] ICCU, ERPANET, *Normative e linee d'azione per la conservazione delle memorie digitali. Un'indagine conoscitiva*, a cura di L. Lograno, trad. di F. Marini, Firenze 2003, <https://www.iccu.sbn.it/export/sites/iccu/documenti/normative_it.pdf> (ultima consultazione: 16/03/2020).
[6] M. GUERCIO, *The Italian case: legal framework and good practices for digital preservation*, Fondazione Rinascimento Digitale, 2012, < https://www.lettere.uniroma1.it/sites/default/files/428/FirenzeDicembre2012guercio_paper.pdf > (ultima consultazione: 16/03/2020).

# 3. Metodologia

L'indagine oggetto del presente lavoro ha avuto l'obiettivo di analizzare il tema della conservazione dei documenti informatici in ambito sanitario sia dal punto di vista del produttore sia da quello del conservatore.
A tal fine, l'approccio metodologico utilizzato si articola in 3 fasi principali:
1. indagine esplorativa sui diversi aspetti relativi all'attuazione del FSE, tra cui la conservazione digitale dei documenti sanitari;
2. reperimento e analisi dei manuali di conservazione;
3. indagine conoscitiva mediante colloqui con conservatori accreditati AgID, che si occupano prevalentemente di conservazione di documenti informatici in ambito sanitario.

Le attività svolte sono dettagliate nei sottoparagrafi seguenti.

## *3.1. Indagine esplorativa*

Tra febbraio e novembre 2017 è stata condotta un'indagine esplorativa sullo stato di attuazione del FSE, realizzata mediante la predisposizione di questionari *ad hoc* e volta a determinare un insieme di azioni mirate all'identificazione e allo sviluppo di regole, sistemi e servizi riusabili, al fine di supportare le Regioni e Province Autonome nella realizzazione, in maniera condivisa ed omogenea, dei sistemi di FSE, secondo quanto stabilito dal citato DPCM 178/2015. Sono stati realizzati due questionari miranti, tra le altre cose, a rilevare anche informazioni sullo stato di realizzazione di sistemi di conservazione a norma. Il primo questionario è stato indirizzato ai referenti regionali in materia di sanità con l'obiettivo di rilevare dati utili a fare un'analisi sullo stato generale di attuazione del FSE, mentre il secondo è stato indirizzato ai referenti delle singole strutture sanitarie afferenti al Servizio Sanitario Nazionale (SSN)[7], per garantire un'analisi più puntuale dei servizi avviati e dello stato di informatizzazione a livello locale.

La strutturazione e la definizione dei contenuti sono stati concordati con AgID e il tavolo interregionale Area IT Sanità. I questionari sono stati formalizzati e somministrati mediante la piattaforma web open source Lime Survey[8].

In relazione ai contenuti, i questionari presentano una suddivisione in sezioni specifiche che prevedono diverse tipologie di domande (risposta aperta, risposta chiusa, ecc.) in base alla natura dell'informazione da rilevare. In particolare, il questionario rivolto ai referenti regionali è stato strutturato in sette diverse sezioni con domande specifiche sui più importanti aspetti oggetto della rilevazione: i) Infrastruttura IT per l'erogazione di servizi sanitari; ii) Accesso ai servizi di sanità elettronica; iii) Gestione del consenso per il FSE; iv) Codifica delle informazioni cliniche; v) Sistemi di alimentazione e consultazione del FSE; vi) Gestione documenti digitali e fogli di stile; vii) Conservazione dei documenti digitali.

Il secondo questionario, rivolto ai referenti delle strutture sanitarie è stato suddiviso, invece, in quattro diverse sezioni: i) Anagrafica azienda; ii) Data center; iii) Servizi applicativi per la sanità digitale; iv) Conservazione dei documenti sanitari.

### *3.1.1. Risultati*

Ai fini del presente studio sono stati analizzati i risultati relativi alla conservazione dei documenti informatici di entrambi i questionari, con l'intento di effettuare una ricognizione più dettagliata dei modelli in essere nei contesti regionali, permettendo di verificare le azioni intraprese dalle Regioni e dalle strutture interpellate.

---

[7] I questionari sono stati inviati ai referenti per il settore sanitario di tutte le Regioni e Province Autonome e ai referenti delle strutture sanitarie presenti in un apposito elenco fornito da AgID.
[8] Lime Survey è un software open source per la compilazione di questionari online, <https://www.limesurvey.org/> (ultima consultazione: 16/03/2020).

Per quanto concerne il primo questionario, a fronte dei 21 inviti a compilare il questionario, inviati via e-mail a tutti i referenti regionali, 18 sono stati coloro che hanno effettivamente compilato e sottomesso il questionario. Le risposte ricevute sono state opportunamente elaborate al fine di ricostruire i diversi scenari regionali nelle tematiche relative all'attuazione del FSE. Per quanto concerne la conservazione, analizzando le risposte ricevute, solo 3 tra le 18 Regioni e Province Autonome partecipanti all'indagine[9] hanno predisposto un sistema di conservazione ai sensi del DPCM del 3 dicembre 2013[10]. Tutte e tre le strutture, inoltre, si sono affidate a conservatori esterni accreditati per la predisposizione del sistema di conservazione. È doveroso specificare che i risultati numerici riportati sono relativi alle risposte rilevate fino a novembre 2017 e, pertanto, devono essere riferiti a quel periodo di attuazione del FSE.

È importante notare che tutti i sistemi di conservazione predisposti prevedono il rispetto dei formati previsti nell'allegato 2 del DPCM 3 dicembre 2013[11] per gli oggetti destinati alla conservazione. Inoltre, sono state predisposte da tutti i rispondenti misure idonee a garantire la protezione dei dati conservati in caso di eventi dannosi, ai sensi dell'art. 31 del DL 196/2003 *Codice in materia di protezione dei dati personali*, e modalità idonee a garantite la continuità delle operazioni indispensabili per il servizio e il conseguente ritorno alla normale operatività in caso di evento infausto.

Per ciò che concerne il secondo questionario rivolto alle strutture sanitarie, hanno partecipato 156 strutture su un totale di 207 invitate e fra esse solo 73 hanno indicato di avere un manuale di conservazione a norma, inviandone anche il link per la consultazione. Come si evince dal grafico riportato in Fig. 1, sono poche le Regioni la cui totalità delle strutture sanitarie afferenti ha predisposto un sistema di conservazione a norma ai sensi del citato DPCM del 3 dicembre 2013 ovvero Emilia-Romagna, Friuli Venezia Giulia, Provincia Autonoma di Trento, Valle d'Aosta e Veneto.

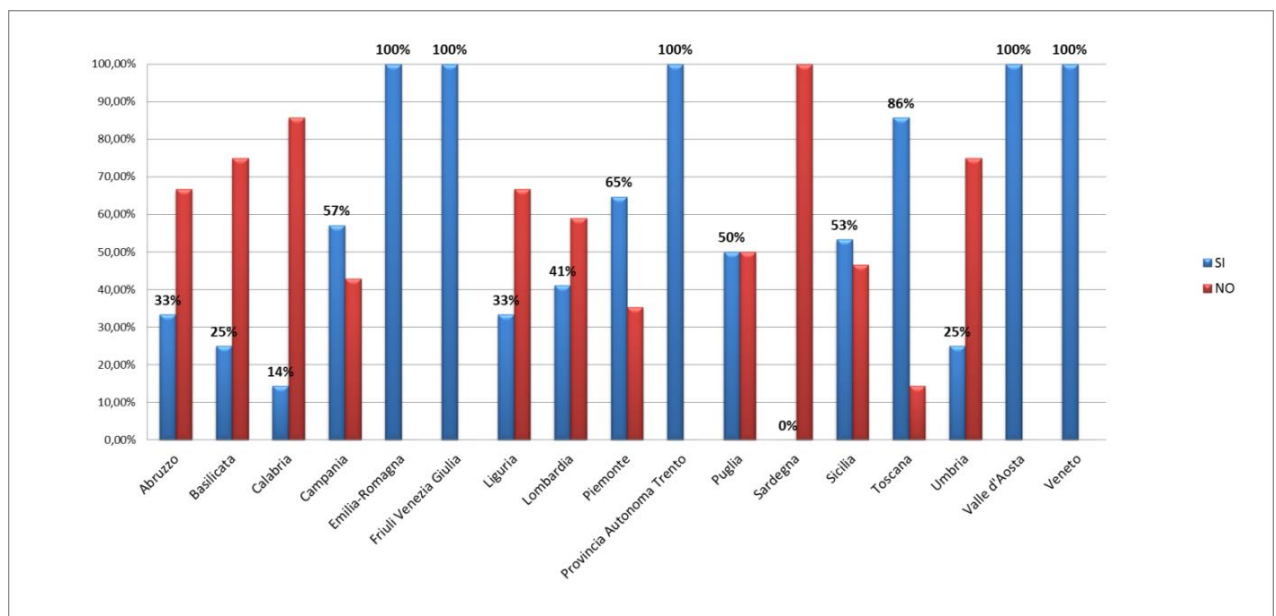

Figura 1 - Percentuali di strutture sanitarie per Regione dotate di un sistema di conservazione conforme alla normativa

---

[9] Provincia Autonoma di Trento, Molise e Valle d'Aosta.
[10] *Regole tecniche in materia di sistema di conservazione, ai sensi degli articoli 20, commi 3 e 5-bis, 23-ter, comma 4, 43, commi 1 e 3, 44, 44-bis e 71, comma 1, del CAD di cui al decreto legislativo n. 82 del 2005*. Consultabile al link: <https://www.gazzettaufficiale.it/eli/id/2014/03/12/14A02098/sg> (ultima consultazione: 16/03/2020).
[11] È importante ricordare che, essendo intervenute nel frattempo le nuove Linee guida AgID, in particolare con l'allegato 2 *Formati di file e riversamento*, le tipologie di formati precedentemente previste sono state ampliate includendo, inoltre, per ognuna di esse, indicazioni sul loro utilizzo legato alla conservazione. Per maggiori dettagli si veda il seguente link: <https://docs.italia.it/AgID/documenti-in-consultazione/lg-documenti-informatici-docs/it/bozza/_downloads/5e42a9d5c8873a1dc532ef5522f5477c/All%202%20-%20Formati%20di%20File%20e%20Riversamento.pdf> (ultima consultazione: 16/03/2020).

Fra le strutture sanitarie rispondenti, solo 5 hanno dichiarato di aver realizzato un sistema di conservazione *inhouse*, tra cui 4 strutture in Regione Lombardia e 1 struttura in Friuli Venezia Giulia.

Per quanto riguarda la predisposizione di uno specifico manuale di conservazione, la distribuzione percentuale fra coloro che lo hanno approntato e coloro che, invece, non lo hanno fatto è piuttosto equilibrata, come si può evincere dai dati riportati in Tab. 1. Alle strutture sanitarie è stato, inoltre, richiesto di indicare specificatamente l'indirizzo web al quale poter reperire il manuale di conservazione.

| Regione | SI | NO | Non risposte |
|---|---|---|---|
| Abruzzo | 33,33% | 33,33% | 33,33% |
| Basilicata | 75,00% | 0,00% | 25,00% |
| Calabria | 14,29% | 57,14% | 28,57% |
| Campania | 28,57% | 57,14% | 14,29% |
| Emilia-Romagna | 100,00% | 0,00% | 0,00% |
| Friuli Venezia Giulia | 100,00% | 0,00% | 0,00% |
| Lazio | | | |
| Liguria | 33,33% | 66,67% | 0,00% |
| Lombardia | 53,85% | 38,46% | 7,69% |
| Marche | | | |
| Molise | | | |
| Piemonte | 76,47% | 11,76% | 11,76% |
| Provincia Autonoma Bolzano | | | |
| Provincia Autonoma Trento | 100,00% | 0,00% | 0,00% |
| Puglia | 60,00% | 40,00% | 0,00% |
| Sardegna | 50,00% | 50,00% | 0,00% |
| Sicilia | 40,00% | 60,00% | 0,00% |
| Toscana | 71,43% | 28,57% | 0,00% |
| Umbria | 0,00% | 25,00% | 75,00% |
| Valle d'Aosta | 0,00% | 100,00% | 0,00% |
| Veneto | 75,00% | 16,67% | 8,33% |

Tabella 1 - Percentuali di strutture sanitarie per Regione che hanno predisposto un manuale di conservazione

Nonostante molte strutture abbiano dichiarato di aver avviato la predisposizione di strumenti e servizi inerenti alla conservazione, non è ancora altrettanto alto il numero di quelle che hanno realizzato un manuale di conservazione. Tuttavia, coloro che conservano lo fanno occupandosi prevalentemente di documenti nativi digitali e rispettando la tipologia di formati previsti nell'allegato 2 del DPCM 3 dicembre 2013.

### 3.2. Reperimento e analisi dei manuali di conservazione delle strutture del SSN

Sulla base dei risultati ottenuti a seguito dell'indagine esplorativa, si è deciso di approfondire lo studio verificando, per il tramite dell'analisi dei manuali di conservazione, l'appropriatezza delle misure adottate dalle strutture sanitarie, per avere un quadro più pertinente sullo stato di attuazione della normativa.

Questa scelta nasce dalla rilevanza che il manuale di conservazione riveste all'interno del sistema di conservazione, poiché ne descrive e ne rispecchia il funzionamento, garantendone, al tempo stesso, l'efficienza e la realizzazione a norma. Esso, infatti, oltre a definire i soggetti coinvolti, i ruoli svolti dagli stessi ed il modello di funzionamento, descrive nel dettaglio le singole fasi del processo di conservazione, le tipologie di oggetti sottoposti a conservazione, le architetture, le infrastrutture utilizzate e le misure di sicurezza adottate. Inoltre, nei casi di affidamento del servizio in *outsourcing* sono al suo interno dettagliati anche i termini del rapporto fra produttore e conservatore accreditato, quali ad esempio definizione delle specifiche operative, delle modalità di descrizione e di versamento nel sistema di conservazione digitale dei documenti informatici e delle aggregazioni documentali informatiche oggetto di conservazione. Ciò detto, il manuale può assumersi quale strumento operativo e descrittivo del modello di conservazione scelto solo se redatto nel pieno rispetto della normativa di riferimento.

Da una corretta lettura del DPCM del 3 dicembre 2013[12], inoltre, si evince che la predisposizione del manuale è di competenza del Responsabile della Conservazione (RDC) interno alla singola struttura sanitaria. Quest'ultimo ha un ruolo centrale nell'ambito di tutto il sistema di conservazione, poiché ne ha la responsabilità e ha il compito di agire d'intesa con tutte le altre figure di rilievo nell'ambito della Pubblica Amministrazione (quali il Responsabile della Sicurezza e il Responsabile dei Sistemi Informativi), nonché con i fornitori di servizi esterni, come ad esempio il Responsabile del Servizio di Conservazione (RDSC), nel caso in cui si scelga un modello *outsourcing*.

Questa seconda fase dello studio è stata svolta tra maggio 2018 e maggio 2019 ed ha previsto la ricognizione dei manuali delle 73 strutture sanitarie rispondenti alla specifica domanda sulla predisposizione del manuale di conservazione, o mediante la verifica dei link indicati dai referenti di queste ultime o, nel caso in cui questo non fosse stato possibile, per i motivi indicati in seguito, mediante la consultazione del sito web delle stesse.

Terminata la fase di ricognizione dei manuali, è stata realizzata un'analisi degli stessi, per verificarne la completezza e l'aderenza alla normativa di riferimento e far emergere eventuali lacune e/o punti di forza nella descrizione del processo di conservazione.

*3.2.1. Risultati*

Delle 73 strutture che hanno dichiarato di avere un sistema di conservazione a norma, su 156 rispondenti al questionario, solo 13 hanno indicato un link che rimandava ad un manuale di conservazione effettivamente riferibile alla struttura così per come indicato dalla normativa. Le altre, invece, rimandavano al manuale del conservatore accreditato affidatario del servizio oppure al manuale di gestione documentale, non considerabili, quindi, ai fini dell'analisi.

Come accennato, il numero esiguo di manuali recuperati attraverso i link forniti dalle strutture sanitarie ha posto la necessità di ampliare tale ricognizione, verificando l'eventuale disponibilità e reperibilità di ulteriori manuali di conservazione, consultando i siti istituzionali delle 134 strutture sanitarie rimanenti, numero dato dalla somma tra le strutture che non avevano dichiarato di avere un sistema di conservazione a norma (83 su 156) e le 51 strutture originariamente invitate, ma non partecipanti al questionario (207 totali meno le 156 rispondenti).

Questa estensione dell'attività di ricognizione dei manuali ha portato a reperire soli ulteriori 5 manuali, arrivando quindi ad un totale di 18. Questo dato è già indicativo dello scarso livello di recepimento della normativa, considerando che la redazione di questo documento rappresenta un obbligo di legge, oltre che il primo step logico necessario per predisporre il sistema di conservazione a norma. Da qui, la necessità di fare maggiore chiarezza sulle responsabilità e sugli obblighi del produttore e sulle modalità di predisposizione di un manuale conforme alla normativa, anche in considerazione del fatto che gran parte dei rispondenti ha, difatti, erroneamente indicato come riferimento al proprio manuale il link al manuale del conservatore accreditato.

L'analisi condotta ha, innanzitutto, evidenziato che il manuale risulta un documento avulso dai contesti organizzativi e la sua redazione, sia per le strutture che utilizzano un sistema *inhouse*, sia per quelle che affidano il servizio in *outsourcing*, si traduce spesso in un mero adempimento burocratico piuttosto che nell'assolvimento di un obbligo di legge e di un passaggio necessario per una chiara e precisa organizzazione del servizio. In particolare, ciò che manca spesso, e che poi genera a cascata tutta una serie di conseguenze, è la responsabilità del produttore nella redazione di questo documento. Da quanto si evince dalle nuove Linee guida pubblicate da AgID in materia di conservazione, infatti, le Pubbliche Amministrazioni (PA), sebbene possano far rinvio al manuale del conservatore accreditato per le parti di processo di competenza di quest'ultimo, hanno comunque l'obbligo di descrivere all'interno del proprio manuale le informazioni relative a: tempi di versamento, tipologie documentali, metadati, modalità di trasmissione dei Pacchetti di Versamento (PDV) e tempistiche di selezione e scarto dei propri documenti informatici[13].

---

[12] Dall'applicazione delle nuove Linee guida il DPCM 3 dicembre 2013 sarà abrogato, ad eccezione dell'art. 13 che rimane in vigore fino alla emanazione delle Linee guida di cui all'art. 29 del CAD.

La conservazione ha come fine ultimo quello della reperibilità di documenti e fascicoli, mantenendo le caratteristiche di autenticità, integrità, affidabilità, leggibilità (ai sensi del CAD, art. 44, comma 1-ter), per cui il manuale di conservazione non può limitarsi a dare descrizioni generiche dei processi fondamentali, quale ad esempio quello di produzione e di esibizione del Pacchetto di Distribuzione (PDD), ma deve necessariamente entrare nel merito delle specifiche pratiche di conservazione effettivamente adottate nel singolo ente preso in considerazione[14].

Dall'analisi sono emersi dati e considerazioni di carattere generale che riguardano l'aggiornamento dei manuali, la reperibilità dei relativi allegati, le tipologie documentali inviate in conservazione e la figura del RDC. Innanzitutto, sulla base dell'anno di pubblicazione, gran parte dei manuali redatti dalle strutture risultano datati e non aggiornati alla normativa corrente (si veda la Fig. 2): solo 3 su 18 infatti risultano pubblicati negli ultimi 3 anni.

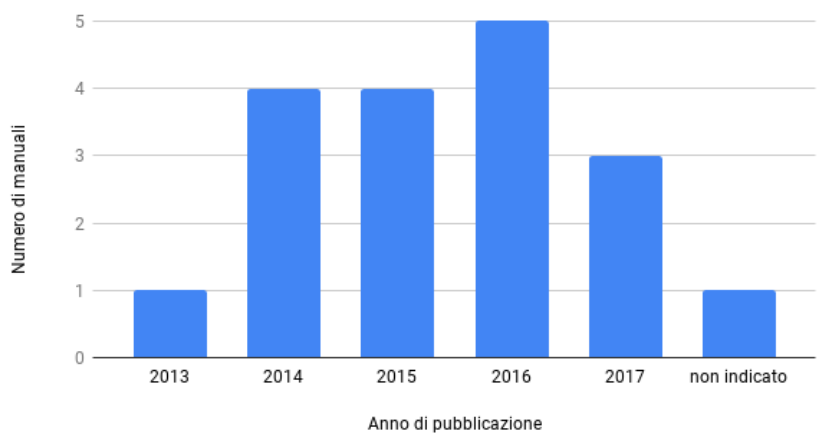

Figura 2 – Manuali di conservazione suddivisi per anno di pubblicazione

Nei casi di conservazione *outsourcing*, inoltre, il mancato aggiornamento ed adeguamento del manuale dell'ente alla normativa corrente fa sì che ci sia incongruenza con il manuale del conservatore accreditato, il quale, invece, assolve l'obbligo di mantenere aggiornate le proprie procedure di conservazione pubblicandone i dettagli sul sito di AgID.

Per quanto concerne il generale approccio alla conservazione, nei 5 casi di conservazione *inhouse*, si nota che nei manuali delle strutture sanitarie spesso non si entra nel merito delle specifiche pratiche di conservazione adottate, ma ci si limita a fornire la descrizione di procedure generiche e di adempimenti prescritti dalla normativa. Il manuale, in altri termini, non assolve alla funzione di rendere pubblici gli aspetti organizzativi così come richiesto dalle regole tecniche. Nei 13 casi di conservazione *outsourcing*, invece, il manuale della struttura risulta maggiormente dettagliato, poichè indica i soggetti coinvolti e i ruoli svolti dagli stessi all'interno della propria struttura e delinea il modello organizzativo di funzionamento dell'attività di conservazione, descrivendo o al suo interno o attraverso un rimando al manuale del conservatore, le caratteristiche tecniche e logiche dei sistemi di conservazione e dei processi effettivamente erogati.

Questa maggiore completezza riscontrata può dipendere dal fatto che il conservatore esterno, perseguendo la sua specifica missione aziendale, mette in campo una serie di competenze tecniche specialistiche di cui non tutte le strutture sanitarie evidentemente dispongono o sono state in grado di dotarsi avendo una *mission* differente, che è appunto quella di cura e assistenza del paziente. Tuttavia, anche la gestione e la corretta conservazione della documentazione sanitaria concorrono a garantire il corretto, efficace ed efficiente percorso di cura del paziente, per far sì che si possa accedere in modo sicuro, veloce e centralizzato alle informazioni cliniche.

---

[13] AGENZIA PER L'ITALIA DIGITALE, *op cit.*, p. 29.
[14] F. DELNERI, *Gli orizzonti della conservazione. Le tre età dell'archivio e il ruolo dei sistemi e degli istituti di conservazione*, in «JLIS.it», January, vol. 10, n. 1, 2019, pp. 12–25.

La mancata reperibilità degli allegati, tecnici e/o contrattuali, a cui negli stessi manuali si fa rimando per la descrizione di aspetti essenziali della conservazione (es. metadatazione, modalità di presa in carico dei PdV, ecc.) rappresenta un limite per la completezza dell'indagine, poiché non restituisce un quadro esaustivo dell'intero processo di conservazione. Dai dati, infatti, emerge che delle 5 strutture che hanno optato per un modello organizzativo *inhouse* solo una struttura rende gli allegati menzionati effettivamente disponibili. Nel caso delle strutture che hanno adottato un modello di conservazione *outsourcing* e che fanno riferimento alla presenza di allegati al manuale della conservazione, l'analisi ha, invece, evidenziato una situazione di sostanziale equilibrio tra le strutture che allegano (in maniera completa o parzialmente) al manuale i documenti citati e quelle che non dotano il manuale di questi documenti complementari né, tantomeno, forniscono indicazioni per il loro reperimento.

I manuali di conservazione redatti dalle strutture, sia in casi di conservazione *inhouse* sia di conservazione *outsourcing*, risultano in alcuni casi riferiti alla conservazione solo di determinate tipologie di oggetti documentali e non della totalità della documentazione, amministrativa e clinica.

Si veda Fig. 3 per i dati relativi ai casi di conservazione *inhouse* e Fig. 4 per quelli di conservazione *outsourcing*.

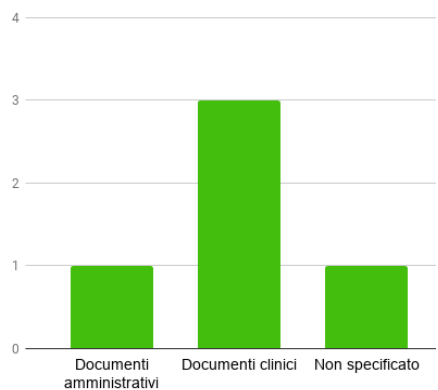

Figura 3 – Tipologie documentali conservate nelle strutture con modello organizzativo inhouse

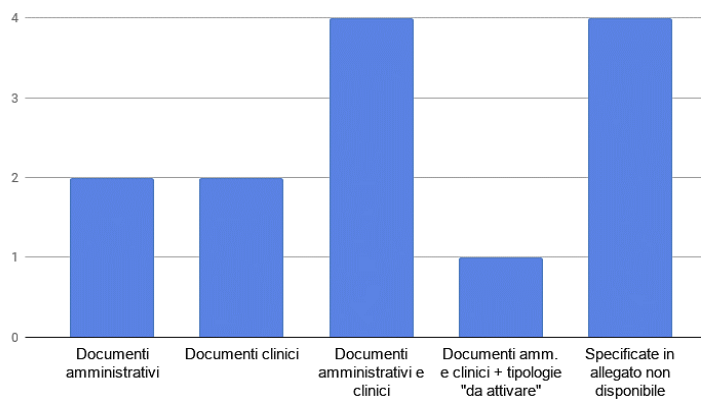

Figura 4 - Tipologie documentali conservate nelle strutture con modello organizzativo *outsourcing*

L'analisi dei manuali ha evidenziato quanto questi siano carenti su aspetti importanti e basilari, quali le generalità e i compiti ricoperti all'interno della struttura organizzativa da parte del RDC, il quale, come ampiamente chiarito, ha, da norma, un ruolo centrale nell'ambito di tutto il sistema di conservazione. A questo proposito, è emerso, infatti, che, mentre la maggior parte dei manuali (ben 14 su 18) specifica correttamente il nome e i compiti del RDC, i restanti 4 manuali analizzati specificano il nome e i compiti del RDSC, che ha compiti e responsabilità complementari, ma non assolutamente sostituibili a quelli di stretta pertinenza del RDC.

### 3.3. Indagine conoscitiva mediante colloqui con conservatori accreditati AgID

I risultati delle attività finora descritte hanno permesso di evidenziare l'ampia diffusione del modello di conservazione *outsourcing*. La scelta di affidare all'esterno il servizio di conservazione è da attribuire ai risultati di una serie di valutazioni costi/benefici legate alla disponibilità di risorse economiche e di strutture adeguatamente attrezzate, nonché di personale con competenze specifiche. Si tratta, infatti, di realizzare un sistema complesso, il cui funzionamento richiede, oltre al costante allineamento alla normativa vigente, l'impiego di figure professionali specializzate e un lavoro di sinergia fra gli attori coinvolti.

Considerando, quindi, l'ampio numero di strutture che ha deciso di affidare il proprio sistema di conservazione in *outsourcing*, la terza fase dello studio si è focalizzata sulla realizzazione di interviste dirette a fornitori del servizio, individuati da AgID come quelli maggiormente attivi nel trattamento di documentazione sanitaria, con l'obiettivo di delineare lo stato dell'arte relativo all'interazione tra produttore e conservatore.

In preparazione ai colloqui de visu con i conservatori, sono state formulate specifiche domande che potessero consentire di indagare tutti gli aspetti fondamentali del processo di conservazione.

Nello specifico, nelle interviste sono stati affrontati tutti i principali aspetti legati alla conservazione digitale, quindi anche quelli di competenza del produttore, proprio per valutare il punto di vista del conservatore accreditato sull'intero processo di conservazione e sull'interazione produttore-conservatore.

*3.3.1. Risultati*

Le criticità rilevate a seguito della fase di ricognizione e di analisi dei manuali hanno trovato ulteriore conferma in quanto emerso durante i colloqui conoscitivi realizzati con alcuni conservatori accreditati.

I conservatori intervistati hanno, in generale, affermato che sono ancora poche le strutture che instaurano un rapporto di reciproca collaborazione e controllo come richiederebbe l'esternalizzazione di un servizio che non ha come fine ultimo quello di deresponsabilizzare, per compiti e competenze, i soggetti interni all'organizzazione affidataria. Essi hanno in altri termini confermato che la maggior parte degli enti demandano il servizio e le decisioni ad esso connesse al conservatore esterno richiedendo altresì supporto nella redazione del manuale per mancanza di competenze specialistiche interne di tipo normativo, archivistico e tecnologico. La carenza di specifiche competenze ha un impatto diretto ed evidente sull'assunzione di ruoli e responsabilità da parte del produttore e del conservatore rispetto alla gestione del ciclo di vita del documento, caratterizzato da specifiche prassi e da strumenti archivistici che devono essere adottati del produttore (ad esempio, la definizione del piano di classificazione e del piano di conservazione)[15].

Le considerazioni più interessanti derivanti dalle interviste riguardano i seguenti argomenti: predisposizione del manuale di conservazione; modellazione dei PDV; metadatazione; definizione dei tempi di versamento e scarto; gestione della fase di migrazione. Tuttavia, pur essendo questi argomenti parte dei principali aspetti caratterizzanti il servizio di conservazione, sono da normativa, ad esclusione dell'ultimo citato, di competenza del produttore dei documenti e non del conservatore. Ciò ad ulteriore testimonianza dell'elevato grado di delega che si tende a fare di questi servizi quando si sceglie il modello di conservazione *outsourcing*.

Per quanto riguarda la predisposizione del manuale di conservazione, emerge nuovamente la difficoltà da parte delle strutture nel redigere autonomamente e correttamente questo documento. I conservatori intervistati hanno, infatti, riferito che la maggior parte dei produttori richiede supporto nella redazione dello stesso, il più delle volte per mancanza di competenze specialistiche interne all'ente, seppure la normativa lo annoveri fra i primi e principali compiti del RDC. Ma, proprio a proposito della definizione di ruoli e responsabilità, secondo quanto riportato dai conservatori intervistati, la maggior parte degli enti non ha individuato all'interno della propria organizzazione le figure professionali con specifiche competenze in materia di gestione e conservazione digitale, così come richiesto dal CAD, oltre a non aver aggiornato il manuale di gestione documentale, laddove esista, in prospettiva della conservazione, rendendo di fatto le due attività come legate e conseguenti.

Per questi motivi, nella maggior parte dei casi, è spesso il conservatore che deve svolgere una fase di verifica preventiva, anche se non di sua competenza, alla formazione dei PDV, per assicurarsi che vi sia la sussistenza dei requisiti minimi necessari per poter inviare i documenti in conservazione. È chiaro come questo passaggio esuli dai compiti e dalle responsabilità del conservatore, poiché coinvolge attività di pertinenza del produttore, partendo dalle regole di versamento fino alla valorizzazione dei metadati obbligatori e, in alcuni casi, passando anche attraverso controlli sul contenuto dei documenti così da intervenire in una prodromica fase di formazione degli stessi. In particolare, secondo quanto riportato dai conservatori intervistati, la maggior parte dei produttori invia in conservazione documenti carenti dal punto di vista dei metadati di struttura, al punto che quasi mai è possibile risalire dagli stessi all'autore, alle responsabilità amministrative e alle informazioni sul contesto tecnologico. Ciò comporta che il più delle volte il conservatore si trovi costretto a dover intervenire in attività di competenza e di responsabilità del produttore, occupandosi di fornire una metadatazione aggiuntiva e di colmare anche la carenza di informazioni su aspetti di natura archivistica. Tale prassi rischia di compromettere l'identificazione univoca e il corretto trasferimento dei documenti in conservazione, unitamente ad una serie di informazioni significative che li rendono individuabili e accessibili per la futura fruizione[16]. I metadati sono, difatti, una componente fondamentale degli archivi digitali e la loro corretta valorizzazione rende possibile la gestione e la rappresentazione del contenuto nei sistemi di conservazione digitali a lungo termine[17]. Per tali ragioni, è emersa la comune istanza di incentivare e supportare la metadatazione minima e procedere alla definizione a livello normativo di metadati aggiuntivi rispetto a quelli minimi, essenziali per il recupero degli oggetti digitali[18] nel tempo, anche in ambienti tecnologici diversi da quello originario.

---

[15] L'art. 68 del Decreto del Presidente della Repubblica n. 445 del 2000, recante *Disposizioni legislative in materia di documentazione amministrativa*, prevede espressamente per le pubbliche amministrazioni l'adozione di un «piano di conservazione degli archivi, integrato con il sistema di classificazione, per la definizione dei criteri di organizzazione dell'archivio, di selezione periodica e di conservazione dei documenti».

Dai colloqui sono emerse, inoltre, non poche incertezze riferibili alla fase di formazione e gestione dei documenti, e alle sue ripercussioni sulla conservazione, in particolare per ciò che attiene al rapporto tra la metadatazione[19] e le politiche di scarto. Durante la fase di creazione dei documenti è, infatti, necessario dare spazio ad una serie di valutazioni archivistiche, mediante la definizione di strumenti *ad hoc*, affinché la gestione e la conservazione dei documenti siano realizzate in maniera corretta. Si fa riferimento, nello specifico, ad un modello di classificazione dei documenti sanitari e ad un adeguato set di metadati, in grado di veicolare, unitamente al documento informatico, e suoi allegati, informazioni sui tempi di conservazione e scarto in relazione alle diverse tipologie documentali.

Infine, durante i colloqui è emerso il delicato argomento della gestione della fase di migrazione, che ha l'obiettivo di garantire all'ente produttore la possibilità di gestire il proprio patrimonio documentale qualora decida di interrompere il contratto di servizio con il fornitore, di gestirlo *inhouse* o di rivolgersi ad altro fornitore. Da quanto riportato dai conservatori, questa fase risulta spesso complessa e farraginosa. Il motivo è da rintracciarsi ancora una volta nella mancanza di indicazioni operative precise su alcuni aspetti legati all'interoperabilità tra sistemi proprietari con caratteristiche tecnologiche differenti[20].

---

[16] S. VITALI, *Passato digitale*, Bruno Mondadori, Milano 2004.
[17] A. ROVELLA, *Metadata consistency and coherence in the digital management and preservation process of administrative records*, in «AIDA informazioni», gennaio/giugno, Anno 37, n. 1–2, 2019, pp. 75–97.
[18] Per una definizione di oggetto digitale si veda, ad esempio, quella della California Digital Library: «an entity in which one or more content files and their corresponding metadata are united, physically and/or logically, through the use of a digital wrapper». <https://cdlib.org/resources/technologists/glossary-of-digital-library-terms/> (ultima consultazione 16/03/2020).
[19] Al momento della formazione del documento informatico immodificabile, devono essere generati e associati permanentemente ad esso i relativi metadati, così come definito nelle Linee guida AgID, le quali definiscono l'insieme dei metadati del documento informatico nell'allegato 5 *Metadati*: <https://docs.italia.it/AgID/documenti-in-consultazione/lg-documenti-informatici-docs/it/bozza/_downloads/68ba1a216597dd078bef95b520f86f14/All%205%20-I%20Metadati.pdf > (ultima consultazione: 16/03/2020).
[20] È pur vero che è attivo, a tal proposito, il Gruppo di Lavoro *Elaborazione di aggiornamenti dello standard UNI SInCRO*, nell'ambito dei lavori del Forum della conservazione, che cura degli aspetti relativi all'interoperabilità. Così come anche specificato nelle già citate Linee guida AgID di recente pubblicazione, l'interoperabilità tra i sistemi di conservazione dei soggetti che svolgono attività di conservazione è garantita dall'applicazione delle specifiche tecniche del pacchetto di archiviazione definite dalla norma UNI 11386:2010 Standard SInCRO (UNI SInCRO) - Supporto all'Interoperabilità nella Conservazione e nel Recupero degli Oggetti digitali, nel quale viene individuato uno schema XML, una struttura dati ampiamente condivisa da adottare per supportare l'interoperabilità tra i sistemi in caso di migrazione. Tale standard è andato a colmare il vuoto che esisteva nell'ambito della conservazione sostitutiva, dove non era possibile una verifica della corretta conservazione tra due sistemi di conservazione che non utilizzavano lo stesso software, o che usavano versioni differenti dello stesso software. Attualmente, è in fase di consultazione l'aggiornamento dello standard, che sarà pubblicato tra qualche mese e che prevede una revisione dello schema XML, con integrazione di nuovi elementi, oltre ad includere chiarimenti concettuali rispetto alla versione del 2010.

## 4. Discussione e conclusioni

Il presente contributo ha inteso fornire una panoramica sullo stato di attuazione in Italia delle pratiche di conservazione dei documenti digitali, con particolare riferimento a quelli di tipo sanitario, al fine di fare emergere criticità utili a definire i prerequisiti di un corretto approccio al processo di conservazione. Lo studio ha previsto un'indagine esplorativa destinata alle Regioni e Provincie Autonome e alle singole strutture sanitarie nazionali, l'analisi dei manuali di conservazione di alcune di queste e un'indagine conoscitiva mediante colloqui diretti con alcuni conservatori accreditati. Mediante queste tre fasi di analisi, si è cercato, quindi, di identificare quelle aree strategiche nella predisposizione e gestione di un sistema di conservazione a norma, che necessitano di interventi a livello tecnico e normativo, al fine di riuscire ad offrire una spinta positiva al settore attraverso una serie di indispensabili indicazioni operative.

I risultati evidenziati dall'analisi convergono sulla necessità di intervenire sulle specifiche responsabilità del produttore dei documenti, relative alle pratiche di gestione documentale e propedeutiche alla conservazione.

Sono emersi, infatti, temi di particolare interesse che riguardano l'integrazione fra sistemi di gestione documentale e sistemi di conservazione che, per quanto distinti, dovrebbero essere strettamente collegati a livello logico. I conservatori hanno, infatti, evidenziato l'impellenza di regolamentare questo aspetto facilitando l'integrazione fra i due sistemi, così da garantire continuità al ciclo di vita del documento, oltre che una maggiore efficacia del processo di conservazione.

L'indagine svolta ha permesso, quindi, di formulare una serie di precondizioni necessarie cui adempiere per un ente che si appresta a realizzare un sistema di conservazione. Sulla base di quanto emerso, è stato, infatti, possibile mettere in luce una lista di prerequisiti cui il produttore dovrebbe attenersi per creare le condizioni necessarie nel passaggio dalla fase di gestione documentale a quella di conservazione e, anche, nell'eventuale passaggio di competenze, in caso di conservazione *outsourcing*, tra il produttore ed il conservatore, così da agevolare i rapporti tra le parti. Esistono, infatti, aspetti importanti che il soggetto produttore deve gestire correttamente secondo quanto concordato con il conservatore scelto ed in linea con la configurazione del sistema di conservazione, così da assicurare la compatibilità fra le policies dichiarate e quelle effettivamente configurate.

I suddetti prerequisiti riguardano principalmente le modalità di formazione dei documenti e la preparazione dei PDV per l'invio in conservazione.

Si rivelano fondamentali, però, anche altri aspetti organizzativi, quali la presenza di personale con competenze informatiche e normative specifiche all'interno dell'ente produttore, e con un'adeguata preparazione archivistica, affinché sia possibile redigere il manuale di conservazione, definire correttamente il titolario o piano di classificazione e il piano di conservazione e di associarli per arrivare a indicare chiaramente: tipologie documentali da portare in conservazione; tempistiche di versamento; tempi di eliminazione/scarto.

È, poi, essenziale che vengano correttamente creati i PDV da parte del produttore, al fine di prevenire eventuali errori ed il conseguente rifiuto dei pacchetti da parte del sistema di conservazione, garantendo il buon esito delle verifiche di presa in carico, attraverso: la definizione, per ciascuna tipologia documentale, di formati adeguati e ammessi dalla normativa di riferimento; l'associazione al documento di un adeguato set di metadati, in grado di veicolare, unitamente al documento informatico, e suoi allegati, informazioni anche su fascicolazione/aggregazione e su tempi di conservazione e scarto in relazione alle diverse tipologie documentali.

Infine, è indispensabile che i produttori possano interagire con i fornitori dei sistemi documentali per poter definire, in maniera anche dinamica ed adattabile alle esigenze aziendali, i metadati da associare ai documenti man a mano che essi vanno formandosi.

Come anticipato nei primi paragrafi, il presente studio, e in particolare l'indagine esplorativa, ha avuto inizio nel 2017, anno in cui era ancora in fase di attuazione l'applicazione della normativa sul FSE. I dati raccolti attraverso l'indagine, anche relativamente agli aspetti legati alla conservazione digitale dei documenti sanitari, hanno, quindi, mostrato una fotografia che potrebbe non corrispondere completamente a quanto riscontrabile oggi effettuando le medesime indagini, seppur gran parte delle problematiche riscontrate siano state evidenziate anche a seguito dei colloqui con i conservatori accreditati. C'è da considerare, inoltre, che nell'intervallo di tempo considerato e sino a novembre 2019 non si avevano ancora a disposizione le nuove disposizioni raccolte nelle Linee guida in materia di conservazione pubblicate da AgID, strumento che chiarisce questioni di primaria importanza, rispetto alle precedenti regole tecniche. Esse sono state pensate e redatte nella prospettiva di formare un unicum normativo in materia, e rappresentano, quindi, un passo in avanti per colmare i gap normativi emersi anche nel corso del presente studio, ma saranno da valutare le ricadute in termini di recepimento da parte degli *stakeholders* della conservazione.

Per superare questi limiti e attualizzare lo studio sarà predisposta, inoltre, una nuova fase di indagine rivolta alle aziende sanitarie, che vada ad indagare la situazione tanto di coloro che sono già attivi in questo ambito, quanto di coloro che sono in fase di organizzazione del sistema di conservazione, così da avere un panorama il più ampio e attuale possibile e ricomprendere tutte le casistiche rilevanti per la definizione di un allegato tecnico per la conservazione dei documenti digitali in ambito sanitario.

In conclusione, si può affermare che la realizzazione di sistemi di conservazione a norma in ambito sanitario risulta ancora poco diffusa e, anche laddove presente, l'aderenza alla normativa vigente non è ancora sufficiente a garantire livelli di qualità elevati per tali servizi. Molti dei gap riscontrati pongono l'accento sui compiti e sulle responsabilità, spesso sottovalutate, del produttore e sull'importanza che dovrebbe assumere un corretto processo di gestione del documento sin dalla sua fase di formazione, poi propedeutica alla conservazione.

La buona riuscita delle politiche di conservazione, ma soprattutto la corretta tenuta, accessibilità e fruibilità nel tempo delle risorse conservate dipendono quindi da molteplici fattori quali l'aderenza alla normativa vigente, la predisposizione e la volontà degli enti produttori e, infine, la collaborazione fra esperti di settore.

## Ringraziamenti

## Riferimenti bibliografici

AGENZIA PER L'ITALIA DIGITALE, *Linee guida sulla formazione, gestione e conservazione dei documenti informatici*, Release bozza 13 febbraio 2020, <https://docs.italia.it/media/pdf/lg-documenti-informatici-docs/bozza/lg-documenti-informatici-docs.pdf>.
Decreto del Presidente del Consiglio dei Ministri 3 dicembre 2013, *Regole Tecniche in Materia di Sistema di Conservazione,* <http://www.AgID.gov.it/sites/default/files/leggi_decreti_direttive/dpcm_3-12-2013_conservazione.pdf>.
Decreto del Presidente del Consiglio dei Ministri 29 settembre 2015, n. 178, *Regolamento in materia di Fascicolo Sanitario Elettronico*, <http://www.gazzettaufficiale.it/eli/id/2015/11/11/15G00192/sg>.
Decreto del Presidente della Repubblica 28 dicembre 2000, n. 445, *"Disposizioni legislative in materia di documentazione amministrativa. (Testo A),* <https://www.camera.it/parlam/leggi/deleghe/00443dla.htm>.
Decreto Legislativo 7 marzo 2005, n. 82, *Codice dell'Amministrazione Digitale*, <http://www.camera.it/parlam/leggi/deleghe/05082dl.htm>.

Decreto Legislativo 26 agosto 2016, n. 179, recante *Modifiche ed Integrazioni al Codice dell'Amministrazione Digitale di cui al decreto legislativo 7 marzo 2005, n. 82, ai sensi dell'articolo 1 della legge 7 agosto 2015, n. 124, in materia di riorganizzazione delle amministrazioni pubbliche*, <http://www.gazzettaufficiale.it/eli/id/2016/09/13/16G00192/sg>.

DELNERI F., *Gli orizzonti della conservazione. Le tre età dell'archivio e il ruolo dei sistemi e degli istituti di conservazione*, in «JLIS.it», January, vol. 10, n. 1, 2019, pp. 12–25.

GUERCIO M., *The Italian case: legal framework and good practices for digital preservation*, Fondazione Rinascimento Digitale, Firenze 2012.

ICCU, ERPANET, *Normative e linee d'azione per la conservazione delle memorie digitali. Un'indagine conoscitiva.*, a cura di L. Lograno, trad. a cura di F. Marini, Firenze 2003.

PIGLIAPOCO S., *La conservazione digitale in Italia. Riflessioni su modelli, criteri e soluzioni*, in «JLIS.it», January, vol. 10, n. 1, 2019, pp. 1–11.

Regolamento (UE) n. 910/2014 del Parlamento europeo e del Consiglio, del 23 luglio 2014*, in materia di identificazione elettronica e servizi fiduciari per le transazioni elettroniche nel mercato interno e che abroga la direttiva 1999/93/CE.*

ROVELLA A., *Metadata consistency and coherence in the digital management and preservation process of administrative records*, in «AIDA informazioni», gennaio/giugno, Anno 37, n. 1–2, 2019, pp. 75–97.

SCOLARI A., PEPE M., MESSINA M., LEOMBRONI C., CIROCCHI G., BERGAMIN G., *Appunti per la definizione di un set di metadati gestionali-amministrativi e strutturali per le risorse digitali*. Gruppo di studio sugli standard e le applicazioni di metadati nei beni culturali promosso dall'ICCU, versione 0, 2002, <https://www.iccu.sbn.it/export/sites/iccu/documenti/MetaAGVZintroduzione.pdf>

VITALI S., *Passato digitale*, Bruno Mondadori, Milano 2004.